# On the mechanical *β* relaxation in glass and its relation to the double-peak phenomenon in impulse excited vibration at high temperatures

Jianbiao Wang[a], Xu Wang[b], Haihui Ruan[a]*

[a] Department of Mechanical Engineering, The Hong Kong Polytechnic University, Hung Hum, Kowloon, Hong Kong, China

[b] Department of Mechanical Engineering, City University of Hong Kong, Kowloon Tong, Kowloon, Hong Kong, China

**Abstract**

A viscoelastic model is established to reveal the relation between *α*-*β* relaxation of glass and the double-peak phenomenon in the experiments of impulse excited vibration. In the modelling, the normal mode analysis (NMA) of potential energy landscape (PEL) picture is employed to describe mechanical *α* and *β* relaxations in a glassy material. The model indicates that a small *β* relaxation can lead to an apparent double-peak phenomenon resulted from the free vibration of a glass beam when the frequency of *β* relaxation peak is close to the natural frequency of specimen. The theoretical prediction is validated by the acoustic spectrum of a fluorosilicate glass beam excited by a mid-span impulse. Furthermore, the experimental results indicate a negative temperature-dependence of the frequency of *β* relaxation in the fluorosilicate glass S-FSL5 which can be explained based on the physical picture of fragmented oxide-network patches in liquid-like regions.

## 1. Introduction

Properties of glass vary with the time due to the internal structural relaxation driven by thermal fluctuation [1]. Experiments have shown that the relaxation happens in almost all the timescales which can be roughly separated into three types when it is close to glass transition temperature ($T_g$) [2, 3]: (1) The primary one, named *α*-relaxation with the typical timescale >$10^{-3}$s, is associated with structural relaxation and play the main role in glass transition; (2) the secondary relaxation with the timescale of

---

$10^{-8} \sim 10^{-3}$s, which is often called (slow) $\beta$ relaxation, is related to localized atomic motion though a mechanism that is still vague; (3) and the third relaxation with the timescale of $10^{-8} \sim 10^{-12}$ s, usually called fast $\beta$ relaxation, could be related to the rattling motion of caged particles [4]. At present, lots of efforts have been devoted to the understanding of $\beta$ relaxation [3, 5-12] because it helps disclose the nature of glass transition [10] and adjust the properties of glass [13-15].

To reveal $\beta$ relaxation, many experimental methods [3, 5-9] have been applied. Among others, dielectric spectroscopy (DS) is the most effective because of the wide frequency range [7, 8] that a DS can swap. Johari and Goldstein [16, 17] first revealed $\beta$ relaxation in several molecular glasses by DS in 1970. Successively, $\beta$ relaxation has been found in the dielectric spectra of polymers [18] and other molecular glass [19]. Based on the measurements of DS, $\beta$ relaxation could further be categorized into two types [20]: a separated secondary relaxation peak or an excess wing of the $\alpha$-relaxation peak. The temperature scaling law of the two manifestations of $\beta$ relaxation seems disparate below $T_g$, that is the average relaxation time of a separated $\beta$ peak strictly follows an Arrhenius behavior [20], whereas that of an excess wing follows a super-Arrhenius law (for example, a Vogel–Fulcher–Tammann (VFT) law) that is strongly coupled to the corresponding $\alpha$-relaxation. When the temperature is close to or higher than $T_g$, the characteristic time of $\beta$ relaxation of some glass formers increases with temperature, which disagrees with the intuition that higher temperature leads to shorter relaxation time [7, 21-23]. This counterintuitive relation has also been found in a water-absorbed porous glass [12], suggesting an intricate mechanism [21] that is still unclear.

The mechanical response of glass has also been employed to study $\beta$ relaxation in it, especially in the cases that DS is unsuitable, for example, metallic glasses [10]. Moreover, mechanical measurements could reveal more internal dynamics than DS because the stress relaxation of a glassy material is related to all diffusion modes, whereas the dielectric response was only related to the reorientation of dipoles [9]. For example, the rotational diffusion about the $C_{2v}$ axis in poly(methyl methacrylate) (PMMA) does not induce the change of dielectric properties; therefore, only mechanical approach can reveal the corresponding relaxation process [9]. For metallic glasses, the internal friction associated with $\beta$ relaxation can only be observed through mechanical means [11] because there is no re-orientation of atomic dipoles. Johari [24] suggested that a mechanical $\beta$ relaxation is essentially due to the translational motion of atoms in metallic and other glasses, which is consistent with the conception of "islands of mobility" proposed by Johari and Goldstein [16].

However, the mechanical approach is much less used to detect the characteristic frequency of $\beta$ relaxation because of the difficulties to achieve a measurement with a wide frequency range. When the test frequency is lower than $10^3$ Hz, some forced vibration methods, for example, dynamic thermomechanical analysis (DMA), can be employed. When the test frequency is larger than $10^9$ Hz, some scatting methods, for example, inelastic light scatting, can be adopted [25]. But for the frequency between these two regimes, there is no standard approaches or commercialized facilities. To expand the frequency range in mechanical tests, Hecksher et al. [26] fused seven different methods with their self-developed facilities. In addition, the decayed free vibration based on the impulse excitation technique (IET) [27] can also be used to study the relaxation behaviors of glassy materials [15, 28, 29]. It should be noted that IET is based on the free vibration of samples, whereas DMA and the approach adopted by Hecksher et al. [26] are based on forced vibration. Comparing with forced-vibration approaches, IET cannot achieve a frequency scan because the natural frequencies of a sample are a series of discrete values. However, the simple and standardized [30, 31] setup of IET, the extended frequency into the ultrasound range ($10^3$ – $10^6$ Hz), and the applicability at a temperature as high as 1750 °C [32] makes it an useful alternative to study relaxation behavior of glasses at the frequency outside the assessable range of DMA. Recently, Liu and Zhang [33] found two adjacent peaks in the acoustic spectrum of a PMMA beam excited by IET, which was ascribed to $\beta$ relaxation. Their experimental results indicate that the Fourier spectrum of an excited beam also contains the information of $\beta$ relaxation.

The dynamic response of a structure subjected to an impulse may reflect the relaxation kinetics inside the material. However, this relation is implicit, which requires a physics-based constitutive model to bridge them. Therefore, in the following, we describe a physical model of $\beta$ relaxation based on the conceptual picture of the potential energy landscape (PEL) and then establish a simplified viscoelastic model based on it. Experimentally, we obtained the double-peaked acoustic spectra of a fluorosilicate glass that validates the $\beta$ relaxation phenomenon predicted by the theoretical model.

**2. Theoretical modeling**

*2.1 The viscoelastic model based on the theory of normal mode analysis*

We follow the normal mode analysis(NMA) to study $\alpha$-$\beta$ relaxation [34]. In this model, $\alpha$ process is caused by the spontaneous hoping among local minima, also called inherent structures (ISs), of the potential energy landscape of glassy material, and $\beta$ process originates from the interaction of atomic oscillations in the basins associated with different ISs. To simplify the analysis, harmonic oscillation is assumed, which can be treated as a combination of the instantaneous normal modes (INM). Both processes lead to the relaxation of physical quantities. For example, Keyes [35] has applied NMA to model the $\alpha$-$\beta$ relaxation of polarizability dynamics in $CS_2$ and achieved a good fit for his atomic simulation. In the present work, we extend the model to describe mechanical $\alpha$-$\beta$ relaxation.

Formally, the constitutive relation of a linear viscoelastic material can be written in the form of hereditary integral:

$$\sigma(t) = E_\infty \varepsilon_0 C(t) + E_\infty \int_0^t C(t-\zeta) \frac{d\varepsilon(\zeta)}{d\zeta} d\zeta, \qquad (1)$$

where $\sigma(t)$ and $\varepsilon(t)$ are the stress and strain at time $t$; $C(t)$ is the relaxation function; $\varepsilon_0$ is the instantaneous strain at $t = 0$. This constitutive relation describes a microscale representative volume element (RVE), which can be divided into many atomic subsystems that can be treated as isolated atomic groups with different ISs. Based on the Green-Kubo relation, the relaxation function is proportional to the stress autocorrelation function (SAF):

$$C(t) \propto \langle \tilde{\sigma}(t)\tilde{\sigma}(0) \rangle, \qquad (2)$$

where $\tilde{\sigma}(t)$ is the instantaneous stress of a subsystem, and "< >" means the average of all atomic subsystems. Following the NMA [35], the fluctuation within a PEL basin is assumed to be harmonic and the stress variation associated with a basin is given as:

$$\tilde{\sigma}(t) \cong \tilde{\sigma}_{IS} + \sum_i \left(\frac{\partial \tilde{\sigma}}{\partial q_i}\right)_{IS} q_i(t), \qquad (3)$$

where $\tilde{\sigma}_{IS}$ is the stress contributed by an IS, $q_i$ is the mass-weighted normal coordinate of the $i$th vibration mode. The SAF can then be expressed as the average of different ISs [35]:

$$\langle \tilde{\sigma}(t)\tilde{\sigma}(0) \rangle = \langle \tilde{\sigma}_{IS}^2 \rangle + k_B T \int \frac{\langle \rho_{IS}(\omega) \rangle}{\omega^2} \cos(\omega t) d\omega, \qquad (4)$$

On the right-hand side of Eq. (4), the first term is the average contribution of ISs, and the second term is the contribution of the average harmonic fluctuations in different ISs. Note that $\rho_{IS}(\omega) = \sum_i \left[ (\partial \tilde{\sigma}/\partial q_i)_{IS}^2 \delta(\omega - \omega_i) \right]$ and $\int_0^{2\pi/\omega_i} (q_i(t))^2 dt = k_B T/\omega_i^2$ have been employed [35, 36].

In Eq. (4), the effect of hopping among ISs is not considered; therefore, $\langle \tilde{\sigma}_{IS}^2 \rangle$ is not a function of time [35] and represents a pure elastic effect. The involvement of structural relaxation brings about memory loss of previous stresses, which can be described by a relaxation function [29, 37]. For simplification, we assume the barrier crossing is an Arrhenius process with a constant barrier height, then an exponential decay can be obtained [36]. Therefore, $\langle \tilde{\sigma}_{IS}^2 \rangle$ should be multiplied by $\exp(-t/\tau_\alpha)$ with $\tau_\alpha$ being the structural relaxation time. This is also because the stress relaxation induced by $\alpha$ relaxation in silicate glass is nearly exponential at the temperature higher than $T_g$ [29]. The second term on the right-hand side of Eq. (4) represents the relaxation induced by harmonic fluctuations, i.e., the relaxation owing to the dephasing induced by the broad distribution of INM frequencies [35]. Though the harmonic term in Eq. (4) is affected by barrier crossing, Cho *et al.* [38] suggested that the additional effect of barrier crossing is not necessary because the dephasing suffices to lead to a reasonable decaying time correlation function. Therefore, the modified SAF involving basin hoping is expressed as:

$$\langle \tilde{\sigma}(t)\tilde{\sigma}(0) \rangle = \langle \tilde{\sigma}_{IS}^2 \rangle \exp(-t/\tau_\alpha) + \int_0^\infty G(\omega)\cos(\omega t)\,d\omega \qquad (5)$$

where $G(\omega) = k_B T \omega^{-2} \langle \rho_{IS}(\omega) \rangle$ is a weighted density of states (WDOS). In principle, $G(\omega)$ can be determined from the eigenvalues of the Hessian matrix of a well-defined atomic model. For example, in the study of the polarity fluctuation of $CS_2$ [35], $\langle \rho_\alpha(\omega) \rangle$ was found to possess several peaks and $G(\omega) \propto \langle \rho_\alpha(\omega) \rangle / \omega^2$ should enhance the contribution of lower-frequency peaks. Following Moore and Space [39], one may assume that $G(\omega)$ describes a bell-shaped distribution, approximated by a Gaussian or Lorentzian function, at a certain frequency range of concern. Assuming that $G(\omega)$ is a Lorentzian function:

$$G(\omega) = \frac{a}{(\omega - \mu)^2 + \gamma^2}, \qquad (6)$$

and submitting it into Eq. (5), the relaxation function is recast as:

$$C(t) = (1-x)\exp(-t/\tau_\alpha) + x\exp(-\gamma t)\cos(\mu t) \qquad (7)$$

where $\mu$ is the central angular frequency of the Lorentzian distribution, $\gamma$ is the half-width at half maximum (HWHM) of a peak, $x = \pi a / (\langle \tilde{\sigma}_{IS}^2 \rangle \gamma + \pi a)$, and $a$ is a constant. Naturally, $x$ can be considered as the proportion of the relaxation contributed by $\beta$ process. It is noted that if $\mu = 0$, Eq. (7) reduces to the scenario of two-step exponential relaxation. In addition, it can be further modified to involve a distribution of relaxation time so that the non-exponential relaxations can also be involved. It is noted that the two-step relaxation function has been used to fit the experimental results of mechanical $\alpha$-$\beta$ relaxation [40, 41]. However, in the following, we shall focus on the case that $\mu$ is nonzero. This makes the relaxation process more complex than a two-step scenario and is indeed necessary to explain our experimental results.

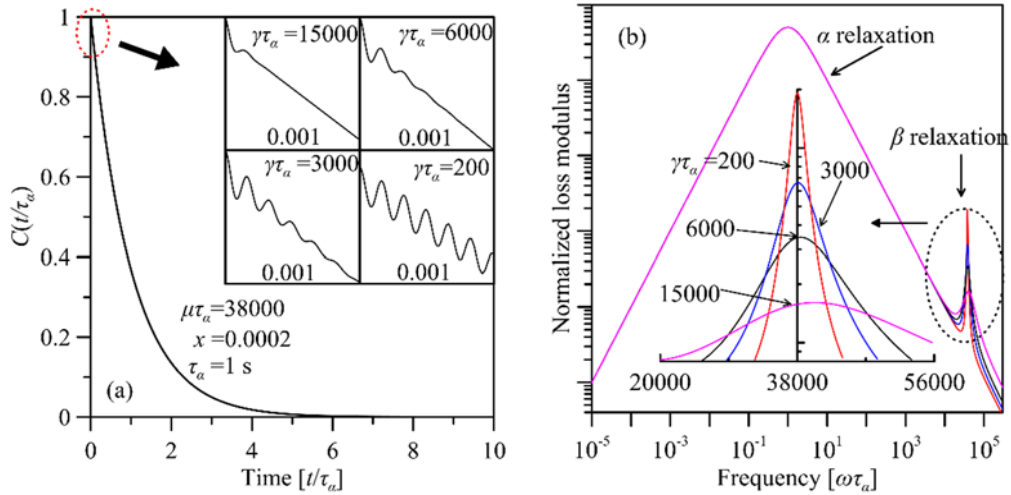

Fig. 1 Examples of stress relaxation (a) and loss spectrum (b) calculated from NMA with Cauchy distribution of $G(\omega)$.

To exemplify the $\alpha$-$\beta$ processes revealed by Eq. (7), we plot the relaxation function in Fig. 1(a) with $x = 0.0002$, $\mu\tau_\alpha=38000$, and amplify the relaxation in initial $0.001\,\tau_\alpha$ in the insets of Fig. 1(a) with different $\gamma$. With the timescale of $\tau_\alpha$, the relaxation function is seemingly a straightforward exponential decay. However, at the very beginning, the stress relaxation could exhibit a plateau if the distribution $G(\omega)$ is very broad ($\gamma\tau_\alpha = 15000$) or oscillate if $G(\omega)$ is sharp ($\gamma\tau_\alpha = 200$~$6000$). It is noted that the initial oscillations are captured in molecular dynamics simulations. For example, based on a bead-spring polymer model, Vladkov and Barrat [42] showed that the short time SAF oscillated and could be fitted with a function identical to the form of Eq. (7). Agrawal *et al.* [43] conducted full-atom molecular dynamics simulations of a polyurea system and clearly showed the transition from initial fast decayed oscillation to long-time decay.

In Fig. 1(b), we covert the stress relaxation into the normalized loss modulus spectrum

$E''(\omega)/E_\infty = \text{Im}[i\omega\tilde{C}(i\omega)]$, where i is the imaginary unit, $\tilde{C}(s)$ is the Laplace transform of $C(t)$ with $s$ being the Laplace variable, and Im[$z$] gives the imaginary part of complex number $z$. It is noted that $\beta$ peaks appear right at the frequency $\omega = \sqrt{\mu^2 + \gamma^2} \approx \mu$ with the width determined by $\sim 2\gamma$. In the scenario of two-step relaxation($\mu$=0), the frequency of $\beta$ relaxation peaks at $\gamma$ [40, 41]. For a mechanical $\beta$ relaxation, this small $\beta$ peak could locate at a frequency range not accessible by a conventional DMA system and/or too small (due to small $x$) to be discernible considering experimental accuracy. Therefore, we introduce the IET experiment and apply the established viscoelasticity model of $\alpha$-$\beta$ relaxation to the response function of excited vibration to examine the possible result of $\beta$ relaxation.

2.2 *Theoretical results of the vibration spectrum of a free-standing beam with α-β relaxation*

Based on the Euler-Bernoulli beam theory, the response function of a free-standing beam can be expressed as[29]:

$$\Gamma_n(s) = \frac{A}{H(s)I(\lambda_n/L)^4 + \rho s^2}, \tag{8}$$

where $H(s) = sE_\infty \tilde{C}(s)$ is the dynamic Young's modulus in the Laplacian domain, $A$ is a variable scaling with the impulse, $I$ is the second moment of inertia of the beam's cross-section, $\lambda_n$ is the modal constant for the $n$th flexural vibration mode, $L$ is the length of the beam, and $\rho$ is the linear density. Substituting $s = i\omega$ into Eq. (8), the Fourier spectrum of the beam vibration can be obtained as:

$$F(\omega) = A \left| \frac{N(s)}{M(s)} \right|^2_{s=i\omega} \tag{9}$$

with

$N(s) = (s + \tau_\alpha^{-1})(s + \gamma + i\mu)(s + \gamma - i\mu)$, and

$M(s) = s(s^2 + s\tau_\alpha^{-1} + \omega_0^2 - x\omega_0^2)[(s+\gamma)^2 + \mu^2] + xs(s+\gamma)(s+\tau_\alpha^{-1})\omega_0^2$,

where

$$\omega_0 = \sqrt{\frac{\lambda_n^4 E_\infty I_z}{\rho L^4}} \tag{10}$$

is the natural frequency. $M(s)$ is a quartic function with four roots. If these roots are all complex, i.e.,

$$\begin{cases} s_{1,2} = -k_1 \pm \omega_{d1} \\ s_{3,4} = -k_2 \pm \omega_{d2} \end{cases}, \tag{11}$$

the Fourier spectrum $F(\omega)$ have double peaks near $\omega_j$ with $\omega_j = \sqrt{\omega_{dj}^2 + k_j^2}$ ($j$=1, 2). To demonstrate, Fig. 2 shows the theoretical double-peaked spectra based on the material parameters used in Fig. 1. and the natural frequency $\omega_0$ in the range of 0.98$\mu$ to 1.02$\mu$. When $\omega_0$ is very close to $\mu$, double peaks are observed and both peak maxima deviate from $\omega_0$. For example, for the cases of $\omega_0$ =0.998$\mu$, $\mu$ and 1.002$\mu$. It is interesting to note that $\mu$ corresponds to the minimum point between the two peaks. This observation can be confirmed based on poles of the reciprocal response function $\Gamma_n^{-1}(s) = M(s)/N(s)$ at $-\gamma \pm i\mu$, which indicates that one of the minima of $F(\omega)$ should be found at $\sqrt{\mu^2 + \gamma^2}$ if double peaks are found. Therefore, the frequency at the minimum point between double peaks can be considered as the frequency of $\beta$ process when $\gamma \ll \mu$. When $\omega_0$ is not so close to $\mu$, as exhibited by the cases of $\omega_0$=0.98$\mu$ and 1.02$\mu$, only one peak can be discerned. It should be noted that even for a single peak case, the frequency at the maximum of the peak could still departure from the natural frequency $\omega_0$ because of the influence of $\beta$ relaxation. This deviation is demonstrated to be about 2% for the cases of $\omega_0$=0.98$\mu$ and 1.02$\mu$. When $\omega_0$ is further deviated from $\mu$, e.g., by changing the dimensions of a sample, the determination of $\omega_0$ using the peak maximum becomes more accurate.

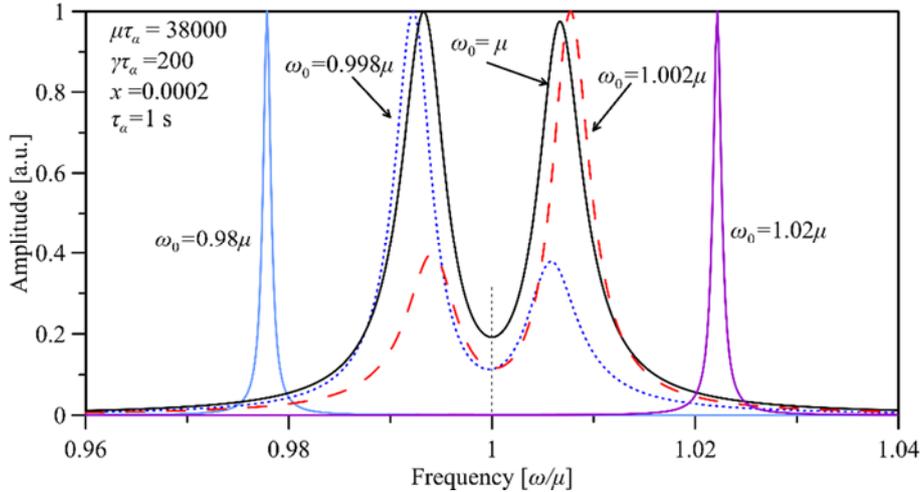

Fig. 2 Examples of double peaks predicted by the viscoelastic model from NMA.

The results shown in Fig. 2 indicate that the excited vibration of a free-standing glassy beam can amplify the $\beta$ process even though it is very subtle in a stress relaxation curve or a loss spectrum, as shown in Fig. 1. However, to capture the double-peaked Fourier spectrum, $\omega_0$ must be very close to $\mu$.

This condition is difficult to meet if $\mu$ is not known *a prior* for a glassy material. In the following, we present a series of clear double-peaked spectrum obtained in examining a fluorosilicate glass.

## 3. Experimental results

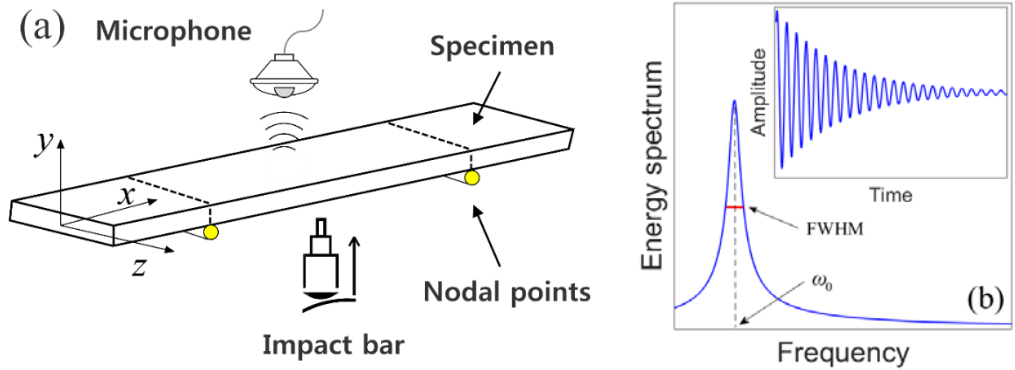

Fig. 3 Sketch of the impulse excitation technique (a) the setup and (b) typical signal and the Fourier transform

In the present work, the IET system HT1600 from IMCE, Belgium, was employed. In the IET stand, a sample is tied by Ø0.2mm PtRh wires, which locates at the two nodal points that have no displacement according to the first flexural vibration mode, as illustrated in Fig. 3(a). The use of the thin metal wire is to approximate to the free-free boundary condition and inhibit higher-order flexural modes. After impacted by a small bar, the sample generates a sound wave corresponding to its decaying vibration, which transmits along a ceramic bar inside the furnace and is sensed by a microphone outside the furnace. The acoustic signal is then transformed to the corresponding Fourier spectrum to analyze the elastic or viscoelastic properties of a material [44], as shown in Fig. 3(b).

The IET setup was employed to examine a fluoride-borosilicate glass S-FSL5 ($SiO_2$(60-70)-$B_2O_3$(10-20)-$F_2$(2-10)-$Al_2O_3$(0-2)-$Sb_2O_3$(0-2), wt.%) procured from OHARA Inc., Japan. The sample exhibited double peaks has the dimensions of 40.08×7.95×1.98 mm³ with mass of 1.5477 g, with the measurement errors <0.01mm and <0.0001g, respectively. S-FSL5 glass has the room-temperature Young's modulus of 62.3 GPa and $T_g$ of 500 °C based on dilatometry measurement. The sample was heated from room temperature to $T_g$+50°C with the prescribed heating rate of 2 °C/min. It was noted that the actual heating rate was 2 °C/min when the temperature was below 547 °C. After that, the heating rate decreased automatically due to the limited controllability of the heating system, *i.e.*, the temperature controller must slowly approach the target temperature for high accuracy and small fluctuations. Consequently, from 549 °C to 550 °C it took 7.5 minutes instead of 0.5 minute. In the tests, only one peak was found when the temperature was lower than $T_g$+10 °C. But after that, a small bump associated with the peak gradually grew to a remarkable secondary peak with the temperature increased, leading the double-peak phenomenon. Fig. 4 shows some typical Fourier spectra. Since the IET system adopted is based on acoustic measurement, the environmental noises were also recorded, leading to small intensity fluctuations (less than 10) at the base of the spectra. These noises does not affect the discernment of a secondary peak. Fig. 4 (a) and (b) shows the cases with a small hump at 511°C and 526°C, respectively, which can be clearly observed after zooming in, as shown in the insets. With the temperature increase, two clear peaks are observed at 531 °C, as shown in Fig. 4(c), whereas only an excess wing associated with the main peak can be found at 540 °C, as shown in Fig. 4(d). In the experiment, these two manifestations appear alternately after 530 °C, which are further exhibited in Fig. 4(e) and Fig. 4(f) for $T$ = 549 °C to 550 °C, respectively. It is noted that the difference between the spectra at 549 °C and 550 °C is substantial, although the temperature difference is only one degree. This could be attributed to the long aging time between the two temperatures, as explained above. The experimental results shown in Fig. 4 has all been well fitted by Eq. (9) using Levenberg-Marquardt arithmetic (The fitting parameters will be discussed later). The adjusted determine coefficient ($R^2$) are also shown in the plots. One may notice that there are still some small humps at both sides of the double peaks as shown in Fig. 4. We call them "shoulder peaks" since they locate on the "shoulders" of the main peaks. The shoulder peaks are owing to the vibration of supporting wires (see the explanation in Appendix).

In Fig. 5, the two frequencies pertaining to the two peak maxima are collected. The lower frequency corresponds to left peak maximum and the higher one corresponds to the right one. All the spectra have been fitted using Eq. (9), which leads to the determination of the natural frequency $\omega_0$ that is also shown in the figure. Obviously, the actual natural frequency $\omega_0$, determined from the instantaneous Young's modulus and dimensions of the beam (Eq. (10)), differs from two apparent frequencies obtained from the two peak maxima. It is noted that the natural frequency is close to the lower frequency, suggesting that $\omega_0 < \mu$. When the temperature is higher than 530°C, the natural frequency has a weaker temperature dependence and departure more from the lower frequency. This slop change of natural frequency around 530 °C suggests there may be some complex structural change in the glass, which needs more investigations.

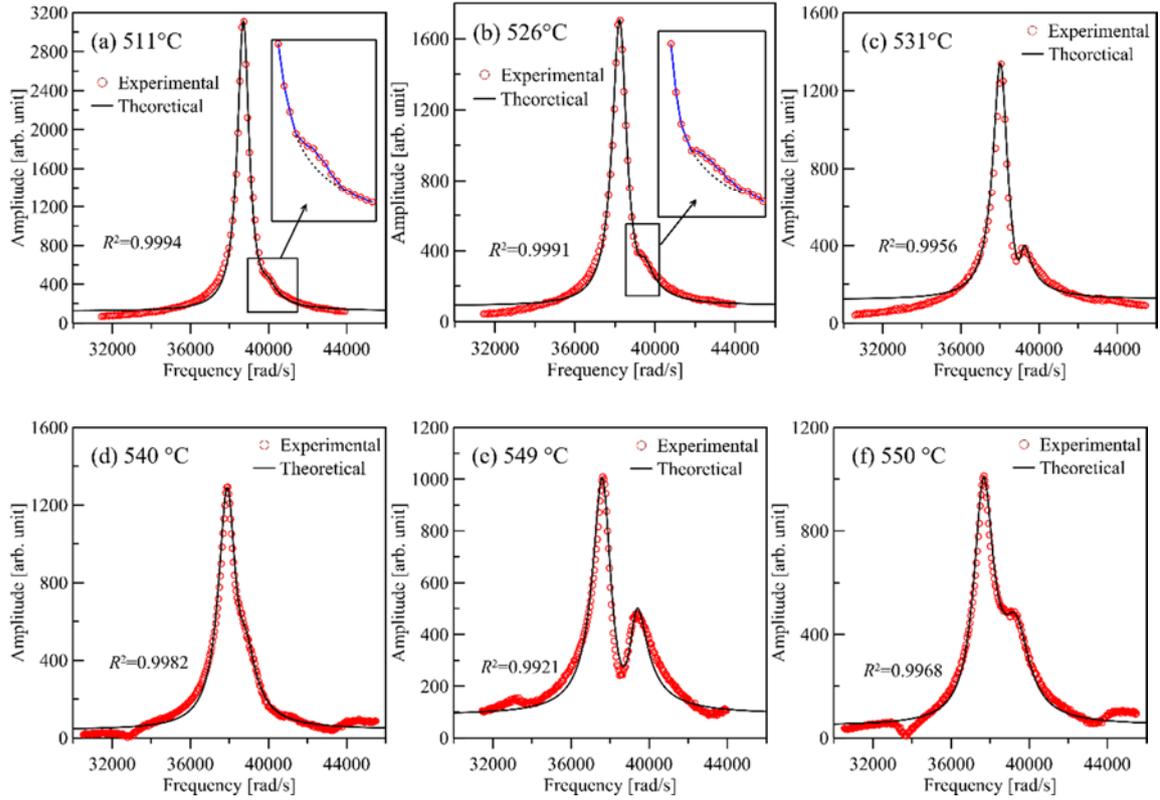

Fig. 4 The double-peak phenomena in Fourier spectra of the free vibration signal of S-FSL5 at different temperature. In the insets of (a) and (b), the solid lines are used to connect the experimental points, while the dashed curves are used to show the tendency of left peak without humps. In the main plot of all figures, the circles are experimental data, and the solid curves are theoretical predictions. $R^2$ is the adjusted determination coefficient representing fitting quality.

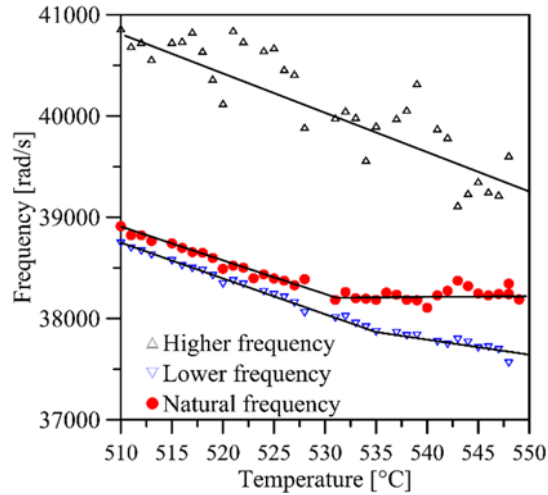

Fig. 5 The temperature dependence of higher/lower frequency from the spectrum, and the calculated natural frequency and relaxation frequency. The lines are artificial trendlines.

## 4. Discussions: temperature dependence of the mechanical $\beta$ relaxation in the fluorosilicate glass

Based on the proposed model, the temperature dependence of $\beta$ relaxation in the fluorosilicate glass is exhibited in Fig. 6. The central frequency $\mu$ and the HWHM $\gamma$ are plotted in Fig. 6(a) and (b), respectively, and the proportion $x$ is plotted in Fig. 6(c). Owing to the experimental noise and also because the proposed model could be still simplistic to describe real physics, all the obtained parameters fluctuates with temperature. However, the general trends of these parameters are clear. It is noted that the frequency associated with the $\beta$ relaxation decreases with temperature, as shown in Fig. 6(a), and that $\gamma$ is weakly dependent on (or slightly decreases with) temperature when $T < 540°C$ and then increase with temperature when $T > 540\ °C$, as shown in Fig. 6(b). This indicates the distribution of INM frequencies becomes broader after 540 °C, which could be ascribed to the increase of the disorderliness of the atomic system. The fraction of $\beta$ relaxation $x$ is smaller than 0.0025, which agrees with previous investigations on the strength of $\beta$ relaxations [16, 17, 45]. In addition, $x$ increases with temperature, especially when the temperature is larger than 530 °C, as shown in Fig. 6(c).

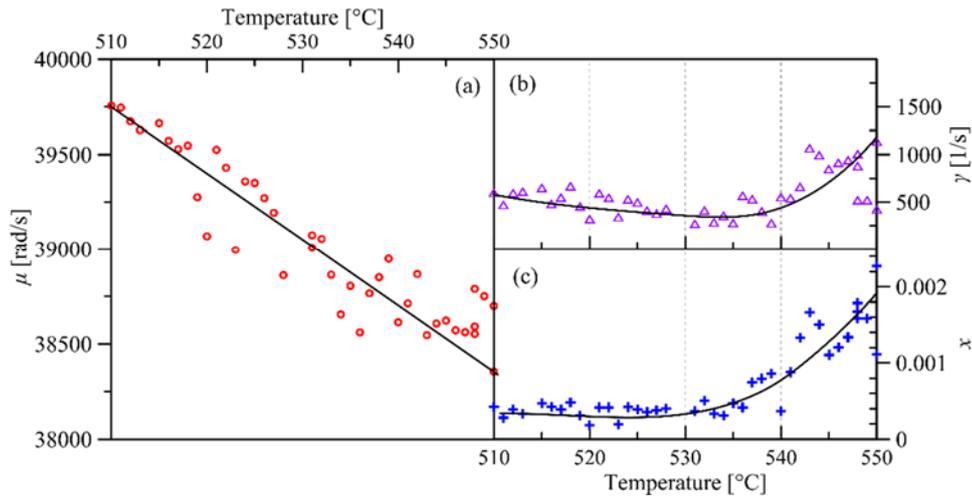

Fig. 6 The temperature dependence of central frequency $\mu$, the HWHM $\gamma$, and proportion $x$. The points are from the theoretical calculation based on experimental data, and the curves are artificial trendlines.

The frequency $\mu$ does not follow Arrhenius or super-Arrhenius law. Such a result is seemingly consistent with the recent experimental work of Hecksher *et. al* [41] who used their self-developed device to reveal that the mechanical $\beta$ peak frequency may also decreases with temperature in squalane. The positive temperature dependence of $\beta$ frequency may be comprehensible [20] if the reciprocal of $\beta$ frequency is considered to be a relaxation time, but the negative temperature dependence of $\beta$ frequency is anomalous. However, the latter is not unusual and also found in DS measurements of various glass formers [7, 12, 21-23]. The existing explanation is phenomenological based on a minimal model (MM) of asymmetric double-well potential [7] or a nonmonotonic relaxation kinetic model (NRKM) [12]. In MM, the two energy wells have different temperature dependence, thus the relaxation time may show anomalous temperature dependence. In NRKM, a rather counterintuitive physical picture is proposed. That is, with temperature increase the total volume of the system changes at a rate smaller than the rate of defect increase. Therefore, the average free volume associated with every defect becomes smaller and then reduce the space of $\beta$ relaxation, leading to the negative temperature dependence of $\beta$ relaxation frequency.

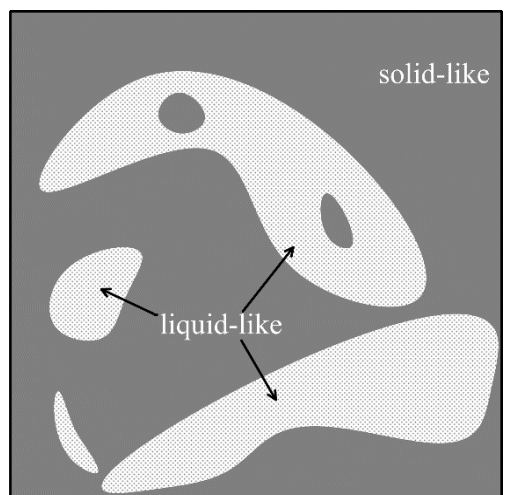

Fig. 7 The sketch of the solid-like and liquid-like region of glass near the glass transition. The grey regions are solid-like and the while regions are liquid-like.

Based on the NMA used in the present work, another view may be provided for comprehending negative temperature dependence of the $\beta$ frequency. Based on our model, the oscillation frequency of $\beta$ process is because the WDOS $G(\omega)$ has a peak at the corresponding frequency range. This requires very weak interactions and large atomic clusters. It is presumed that only in weak bonded regions $\beta$ relaxation could take place [46], thus the negative temperature dependence of $\beta$ relaxation is owing naturally to the weaker interactions when temperature increases and volume expands. In S-FSL5, the atoms are bonded by ionic-covalent interaction in structural polyhedron and connected by the long-range interactions (for example, Coulombic interactions [47]) between polyhedrons. Moreover, the introduction of the network modifier fluorine increases the possibility of isolated polyhedrons. The $\beta$ relaxation is found in the experiment at the temperature higher than $T_g$ but much lower than the melting point. In this temperature regime, the materials experience a transition from solid-like to liquid-like state, which can be described by the picture of Orowan [48], as illustrated in Fig. 7. When the temperature is low, the material must have local mobile regions surrounded by a rigid matrix that did not permit viscous flow. With temperature increase, the sizes and numbers of such regions grow until they are connected, and viscous flow becomes possible. These liquid-like regions are reminiscent of Johari and Goldstein's picture of "islands of mobility" (or "loosely packed isolated regions") [16, 24] which has also been employed by Nemilov [47] to explain $\beta$ relaxation in silicate-based glasses. When the temperature is lower than $T_g$, they provide the room for $\beta$ relaxation of some small atomic clusters. When the temperature is higher than $T_g$, the mobility and size of liquid-like regions increase significantly, and some bigger atomic clusters (mainly oxide-network patches in S-FSL5) fall off from the matrix and take part in the activities of $\beta$ relaxation. Besides, the long-range interactions become also weaker with temperature increase. Therefore, the $\beta$ relaxation can be found at a relatively low frequency which decreases with temperature, as shown in Fig. 6(a). In addition, the fraction of $\beta$ relaxation, namely $x$, should increase with temperature, which is also corroborated by Fig. 6(c).

## 5. Conclusions and remarks

We established a viscoelastic model based on the normal mode analysis of potential energy landscape to describe mechanical $\alpha$ and $\beta$ relaxations in a glassy material. Based on the model, it is predicted that an apparent double-peak phenomenon in the Fourier spectrum of a free beam vibration can be generated by a very weak $\beta$ process when the frequency of $\beta$ relaxation peak is close to the natural frequency of the specimen. This result has been validated by the acoustic spectrum of a fluorosilicate glass (S-FSL5) beam excited by a mid-span impulse. By analyzing the experimental results with the proposed model, it is found that there is a negative temperature-dependence of the $\beta$ frequency in the fluorosilicate glass, which can be explained based on the picture of fragmented oxide-network patches in liquid-like regions.

**CRediT authorship statement**

Jianbiao Wang: Conceptualization, Methodology, Software, Formal analysis, Investigation, Data curation, Writing - original draft. Xu Wang: Investigation, Data curation, Writing - Review & Editing. Haihui Ruan: Conceptualization, Methodology, Validation, Resources, Writing - review & editing, Supervision, Project administration, Funding acquisition.

## Acknowledgment

This work was supported by the Early Career Scheme (ECS) of the Hong Kong Research Grants Council (Grant No. 25200515, Account Code: F-PP27). We are grateful for the support.

**Appendix: On the "shoulder peaks"**

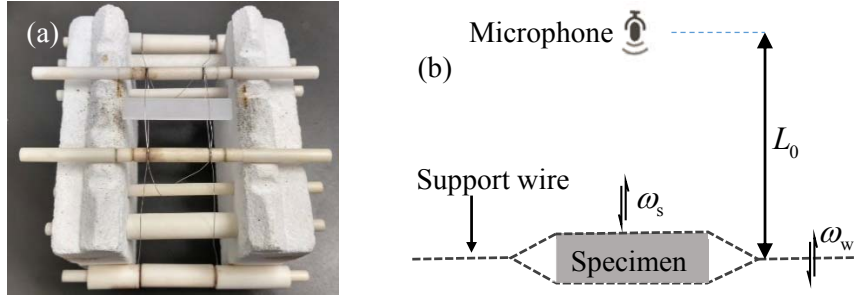

Fig. A1 The vibrations of the detected system

Fig. A1(a) and (b) show how a beam specimen is tied using metal wires in our experiment and the schematic of the testing system in a side view, respectively. After the beam specimen is excited by a tapper, the supporting wires also vibrate, making the beam move up and down and changing the distance between the beam and the microphone. The sound intensity signal collected by the microphone can be expressed as:

$$\Omega_s = \chi \cdot S_s \qquad (A1)$$

where $S_s = A_s \exp(-k_s t)\cos(\omega_s t)$ represents the vibration of a point in the beam sample, $\Omega_s$ is the sound intensity and $\chi$ is the conversion coefficient from the beam displacement to the sound intensity. Considering that in Eq. (A1) $\chi$ is related to the distance $L_0$ between the beam and the microphone, and the vibration of wires, $S_w = A_w \exp(-k_w t)\cos(\omega_w t)$, will slightly change the distance, we thus write $\chi$ as a function of $(S_w+L_0)$. The parameters $S$, $A$, $k$, and $\omega$ used above represent the displacement, amplitude, decay rate, and angular frequency of the specified vibration, respectively, with subscripts "s" and "w" pertaining to specimen and wire, respectively. Since $S_w \ll L_0$, expressing $\chi(S_w + L_0)$ by Tayler's serials at $L_0$ leads to

$$\chi(S_w + L_0) = \chi_0 [1 + \alpha S_w + \cdots] \qquad (A2)$$

where $\chi_0 = \chi(L_0)$ and $\alpha = \chi'(L_0)/\chi(L_0)$. Substituting Eq. (A2) into Eq. (A1) and neglecting the higher-order items, we have

$$\begin{aligned}
\Omega_s &\approx \chi_0 S_s (1+\alpha S_w) \\
&= A_M^* \exp(-k_s t)\cos(\omega_s t) \\
&\quad + A_S^* \exp\left[-(k_s+k_w)t\right]\cos\left[(\omega_s+\omega_w)t\right] \\
&\quad + A_S^* \exp\left[-(k_s+k_w)t\right]\cos\left[(\omega_s-\omega_w)t\right]
\end{aligned} \tag{A3}$$

where $A_M^* = \chi_0 A_s$ and $A_S^* = \frac{1}{2}\chi_0 \alpha A_s A_w$ are the amplitudes of sound signal at the frequencies $\omega_s$ and $\omega_s \pm \omega_w$ respectively, and subscripts $M$ and $S$ pertain to the main peak and the shoulder peaks, respectively.

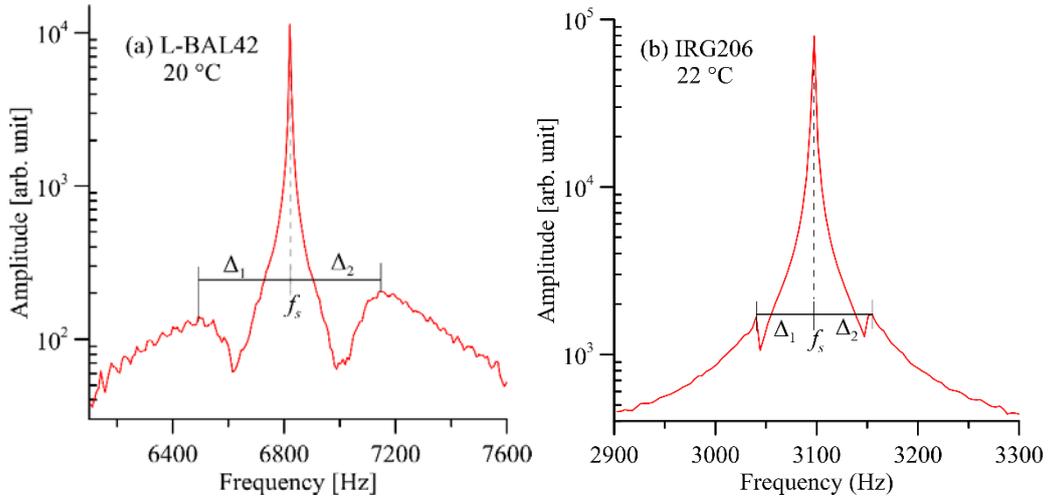

Fig. A2 The Fourier spectrum of the specimen at room temperature. In the plot (a) $f_s$ = 6820.68Hz and $\Delta_1=\Delta_2$=320.44Hz; the Fourier frequency resolutions is 7.63Hz. (b) $f_s$ = 3097.50Hz and $\Delta_1=\Delta_2$=57.22Hz; the frequency resolutions is 3.815Hz

Eq. (A3) indicates that there is a set of symmetric shoulder peaks at $\omega_{1,2}=\omega_s\pm\omega_w$ in the Fourier spectrum of the detected signal. Shoulder peaks are generally very weak and only discernable at a high temperature, at which the main peak of beam vibration would have been severely depressed due to viscous response, as shown in Fig. 4(e) and (f). For a high-quality acoustic signal, shoulder peaks can also be observed even at even at a room temperature. Fig. A2(a) and A2(b) exemplify the Fourier spectra of some glass samples at room temperature (20°C). These glasses are borosilicate (L-BAL42, 40.08×7.98×1.97 mm³, 1.9471g, from OHARA Inc., Japan) and chalcogenide glass (IRG206, 40.03×8.04×2.45mm³, 3.6396g, obtained from Hubei New Hua-Guang Information Materials Co., Ltd, China). Using the logarithmic scale in the ordinate, the shoulder peaks are clearly observed, though they are almost two order lower than the main peak.